\documentclass[11pt]{article}
\begin{document}

%\vspace{0.7in}
\begin{center}\Large{\bf A classical paradox of choice}\\
\vspace{0.4cm}
\normalsize\ J. Finkelstein\footnote[1]{
        Participating Guest, Lawrence Berkeley National Laboratory, 1 Cyclotron Road, Berkeley California 94720\\
        \hspace*{\parindent}\hspace*{.5em}
        Electronic address: JLFINKELSTEIN@lbl.gov}
\end{center}
%\begin{abstract}
%\end{abstract}
%\newpage
\vspace{0.6cm}
In a recent paper ("A quantum Paradox of Choice: More Freedom Makes Summoning a Quantum State Harder"\cite{AK}) Adlam and Kent (henceforth AK) discuss an apparently-paradoxical situation which can arise in the task called "summoning"\cite{K}.  They begin with a parable about a magician who can perform several tricks; when requested to perform any one of them, he can perform the trick which was requested. However, if requested to perform two of his tricks, not only can he not perform both, he cannot perform either one.

It would seem strange that the magician could not choose one of the requested tricks to perform and simply ignore the other request.  AK write 
\begin{quote}
The so-called paradox of choice --- more choice can make consumers less happy --- is a familiar concept in economics.  The magician's paradox, however, is much sharper: more freedom in choosing how to execute a task makes it {\em impossible}.  Nevertheless, strange as it may seem, we show that such a situation can indeed arise when quantum mechanics is combined with classical relativity.
\end{quote}One should not conclude from this that quantum effects are necessary for there to be such a situation, so in this note I point out that such a situation could also arise from purely {\em classical} effects (including classical relativity). 

The apparently-paradoxical situation which AK analyze involves "summoning", and it is only the no-cloning principle of quantum mechanics that prevents {\em all} requests from being fulfilled; I am certainly not suggesting that there is a classical explanation of the restrictions on summoning derived in ref.\ \cite{AK} and \cite{K}.  However, even without the no-cloning restriction or any other non-classical consideration, there could be in principle the following analogue of the AK paradox:
\begin{quote}
There are two agencies, called A and B; B might request that A perform one of several tasks.  If it is agreed that B will not request more than one task, then A can indeed perform the task which was requested.  But if B has the option of requesting more than one task, then A might not be able to perform even one of them.
\end{quote}

Here is a simple example, for which quantum restrictions are not needed: Say there are two laboratories called L and R;  let D be the distance between them, and T = R/c the time required for a signal traveling at the speed of light to go either from L to R or from R to L (all times and distances as measured in the frame in which both L and R are at rest).   There are two tasks which B might request of A:
\begin{quote}
\underline {Task 1} Send a signal from L to arrive at R at time T, but do not send any signal from R to L.  If this task is requested, the request is submitted to A in laboratory L at time t=0.
\end{quote}
\begin{quote}
\underline {Task 2} Send a signal from R to arrive at L at time T, but do not send any signal from L to R.  If this task is requested, the request is submitted to A in laboratory R at time t=0.
\end{quote}
Clearly it would not be possible for {\em both} requests to be fulfilled (just as in the AK example where the no-cloning restriction prevents more than one request from being fulfilled). The situation appears paradoxical because B, when making both requests, would be satisfied if {\em either one} were fulfilled.

For this example, say that agency A employs two agents, agent AL stationed in laboratory L and agent AR in laboratory R, and (if only one task can be requested) can accomplish the requested task  simply by having each of these agents do what is requested of them.  For example, if the request for task 1 is submitted at L, then agent AL sends a light signal to R while agent AR (who did not receive any request) does nothing; thus task 1 is accomplished.  Likewise, task 2 can be accomplished if it is the only one requested. However, suppose that agency B has the option of requesting either one or both tasks, and that A still must accomplish the requested task if only one is requested.  It would still be true that agent AL must send a signal when she receives a request for task 1, because for all she knows that is the {\em only} request made, in which case it must be fulfilled; the relativistic no-signaling principle does not allow her to learn at t=0 whether or not task 2 was requested at R.  Likewise, agent AR must send a signal if she receives a request for task 2.  Therefore if both tasks are requested, both signals will be sent, and so neither task will be accomplished.

\vspace{0.4 cm}
Acknowledgment: I would  like to acknowledge the hospitality of the
Lawrence Berkeley National Laboratory, where this work was done.


\begin{thebibliography}{9}

\bibitem{AK} E. Adlam and A. Kent, arXiv:1509.04226 [quant-ph]
\bibitem{K} A. Kent, Quantum Information Processing {\bf 12}, 1023 (2013)
\end{thebibliography}
\end{document}